\documentstyle[12pt,aaspp4]{article}
\begin{document}

\title{Are Gamma-Ray Bursts in Star Forming Regions?}

\author{Bohdan Paczy\'nski}
\affil{Princeton University Observatory, Princeton, NJ 08544--1001, USA}
\affil{e-mail: bp@astro.princeton.edu}

\begin{abstract}

The optical afterglow of the gamma-ray burst GRB 970508 ($ z = 0.835$)
was a few hundred times more luminous than any supernova.  Therefore,
a name `hypernova' is proposed for the whole GRB/afterglow event.

There is tentative evidence that the GRBs: 970228, 970508, and 970828 
were close to star forming regions.  If this case is strengthened with
future afterglows then the popular model in which GRBs are caused be
merging neutron stars will have to be abandoned, and a model
linking GRBs to cataclysmic deaths of massive stars will be favored.
The presence of X-ray precursors, first detected with Ginga, is easier
to understand within a framework of a `dirty' rather than a `clean'
fireball.  A very energetic explosion of a massive star is likely
to create a dirty fireball, rather than a clean one.

A specific speculative example of such an explosion is proposed,
a microquasar.  Its geometrical structure is similar to the
`failed supernova' of Woosley (1993a): the inner core of a massive,
rapidly rotating star collapses into a $ \sim10 ~ {\rm M_{\odot}} $ Kerr 
black hole with $ \sim 5 \times 10^{54} $ erg of rotational energy, while
the outer core forms a massive disk/torus.  A superstrong 
$ \sim 10^{15} ~ $ G magnetic field is needed to make the object operate 
as a microquasar similar to the Blandford \& Znajek (1977) model.  Such
events must be vary rare, $ 10^4 - 10^5 $ times less common than
ordinary supernovae, if they are to account for the observed GRBs.

\end{abstract}


%\section 1
\section{Introduction}

The recent detection of the afterglows following
some gamma-ray bursts (GRB) detected by BeppoSAX opens up a new era in the 
studies of GRBs.  Many afterglows were detected in X-rays (e.g. Costa et al.
1997), but so far only two were detected optically (970228: 
van Paradijs et al. 1997; 970508: Bond 1997), and just one in radio domain
(970508: Frail et al. 1997).  A major breakthrough was the determination
of the absorption and emission line redshift $ z = 0.835 $ for 970508 
(Metzger et al. 1997a,b).

An afterglow is created by a collision between GRB ejecta and ambient 
medium, be it circum-stellar, interstellar, or intergalactic,
and it is a natural consequence of any relativistic fireball model 
(Rees \& M\'esz\'aros 1992, Paczy\'nski \& Rhoads 1993, Wijers et al. 1997a,
Waxman 1997a,b, Sari 1997, Vietri 1997, and many references therein).
A very impressive confirmation of the
relativistic expansion was provided by the change in the scintillation
pattern of the 970508 radio emission,
as detected by Frail et al. (1997), and predicted by Goodman (1997).

\begin{figure}[t]
\plotfiddle{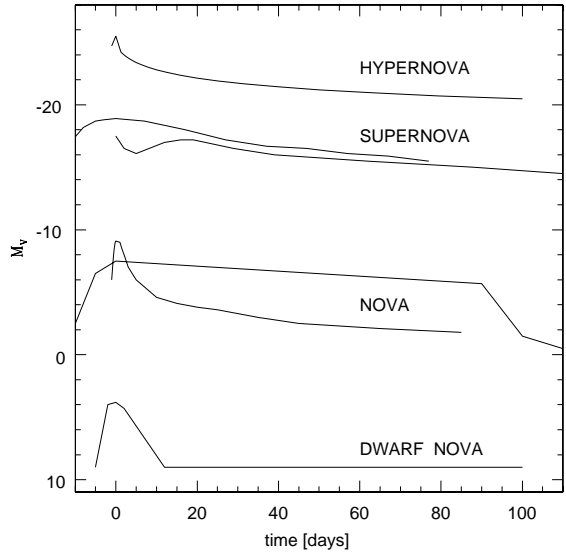}{8cm}{0}{50}{50}{-160}{-90}
\caption{
The absolute visual light curves are shown schematically
for various eruptive variables.  These include one case of a dwarf nova
(U Gem, cf. Fig. 5.23 of Sterken \& Jaschek 1996), two cases of a nova
(V 1500 Cyg and DQ Her, cf. Fig. 40 of Hoffmeister et al. 1985), two
cases of a supernova (SN 1993J, Richmond et al. 1994; SN 1994D, Richmond
et al. 1995), and one case of a hypernova, i.e. the optical afterglow
970508 (Sokolov et al. 1997).
}
\end{figure}

The observed optical emission of 970508 is compared in Fig. 1 with
that of other cataclysmic variables: dwarf novae, novae, and supernovae.
I adopted $ z = 0.835 $ as the redshift of 970508,
and the corresponding K-correction
following Wijers (1997).  With the large range of absolute magnitudes
covered by the light curves an error of $ \sim 1 $ magnitude would be
of little importance.
The optical afterglow 970508 was a few hundred times more luminous than
the brightest supernova.  A few months after the peak it still remains
more luminous than any supernova.  Therefore, it seems appropriate to call it
a hypernova.  I shall use this term to describe the full GRB/afterglow event.
Assuming spherical emission its total energy was $ \sim 10^{52} $ erg
(Waxman 1997b), more than any known supernova.  The terms
like `failed supernova' (Woosley 1993a) or `mini-supernova' (Blinnikov \&
Postnov (1997) seem inappropriate for so energetic explosions.
Note, that hypernova ejecta are relativistic (Goodman 1997, Frail et al. 1997),
making it more violent than any other variable represented in Fig. 1.

The purpose of this paper is to present evidence
that the observed GRBs are related to star forming regions, and
therefore they are not caused by merging neutron stars.
A speculative possibility of the underlying mechanism for the explosion
is outlined: a micro-quasar (Paczy\'nski 1993).  This is a stellar equivalent
of the Blandford \& Znajek (1977) quasar model, powered by a rapid extraction 
of rotational energy of a $ \sim 10 ~ {\rm M_{\odot}} $ Kerr black hole with 
a superstrong $ \sim 10^{15} $ G magnetic field.

%\section 2
\section{Burst locations}

There are only two optical afterglows known, and half a dozen of
non-detections.  In most cases the negative result may be explained
by unfavorable circumstances: a large errors box, a bright moon, or an
inadequate search.  But there was one burst, 970828, for which
the error box was small, the moon was dark, and many deep optical
searches placed stringent upper limits on any optical variable, a factor
$ \sim 300 $ below the expectations based on a simple scaling of the
970228 and 970508 events (Groot et al. 1997,
van Paradijs 1997).  Therefore, the absence
of this optical afterglow has to be taken seriously.
In the following sub-sections all three events are briefly discussed.

\subsection{GRB 970228}

The optical afterglow (van Paradijs et al. 1997, Sahu et al. 1997) has a fuzzy
object next to it.  A recent HST image taken on September 4, 1997. shows
the extended object unchanged, and the point source fading according to the
$ t^{-1} $ law, at a fixed position (Fruchter et al. 1997).  It is
likely that the `fuzz' is a dwarf galaxy at a redshift $ z \sim 1 $.  In
any case, it is not a giant galaxy.  

Note, that by the time a binary neutron star merges it has moved many 
kiloparsecs away from its place of origin, because it had acquired a large 
velocity as a consequence of two consecutive supernovae explosions, even 
if those explosions were spherically symmetric (Tutukov \& Yungelson 1994).
With the velocities of a few hundred $ {\rm km ~ s^{-1} } $ such binaries
escape from dwarf galaxies.  Therefore, if the GRB 970228 was caused
by a merger of two neutron stars, or a neutron star and a stellar mass
black hole, its location near the edge of a dwarf galaxy would be just
a coincidence, though not a very unlikely one.
Future observations of this very faint $ \sim 25 $th mag
extended object may provide some clues.  Its spectrum may show if this
is indeed a galaxy at a moderate redshift, and is it undergoing a vigorous
star formation.

The optical afterglow 970228 makes a weak case against
the merging neutron star scenario, and it is neutral with respect to an
association between the GRB and a star forming region.

\subsection{GRB 970508}

This is the only afterglow for which the redshift has been measured:
$ z = 0.835 $ in absorption (Metzger et al. 1997a)
and in emission (Metzger et al. 1997b).
The [OII] 372.8 nm emission line indicates a normal interstellar medium
rather than AGN.  We also know that it is not resolved by the HST
(Fruchter, Bergeron \& Pian 1997), i.e. the emission line region has
to be very compact.  Therefore, a probability that the positions of
the optical afterglow and the line emission region coincide by chance
is small.  It is reasonable to assume that
the two are related, and that $ z = 0.835 $ is the GRB's redshift,
and not just a lower limit.

The compactness of the emission line object makes it is a good candidate
for a star forming region, and the GRB seems to be associated with it.
This makes it a weak case against a merging neutron star scenario.

\subsection{GRB 970828}

The simplest way to account for the absence of an optical afterglow
has been proposed by Jenkins (1997): extinction by dust.  The case
has been made stronger by Murakami's (1997) report that the ASCA
X-ray spectrum is well fitted by a power law with a low energy absorption
indicating hydrogen column density of $ 4 \times 10^{21} ~ {\rm cm^{-2}} $.
The object is at a high galactic latitude, so the absorption is likely to be
close to the source.  If this burst is at a cosmological distance then the
column density is increased by $ \sim (1+z)^3 $, as the spectral turnover
is at energy $ (1+z) $ times higher than observed.  Also, the observations
made in the R-band correspond to a wavelength $ (1+z) $ times shorter at
the source, with the correspondingly larger interstellar extinction.
Combining all these effects, and adopting a standard
dust to gas ratio may easily provide enough extinction to explain the 
absence of detectable optical afterglow (Groot et al. 1997, van Paradijs 1997).
If this is the correct explanation then GRB 970828 had to be close to
a high density interstellar medium, i.e. close to a star forming region.

Note that gas column density as measured by Murakami (1997) points
to a positional coincidence between the burst and dense interstellar
medium.  So, we have another case for a relation between GRBs and
star forming regions, and against GRBs and merging neutron stars.
Admittedly, the case is weak.  If the GRB is at a very large redshift
then the absorption of soft X-rays may be due to a galaxy which just
happened to be along the line of sight, but is unrelated to the burst.

%\section 3
\section{The burst rate}

In the simplest cosmological scenario for GRB distribution, commonly
accepted till the end of 1996,
it was customary to adopt no evolution: the GRB rate
was assumed to be constant per co-moving volume and co-moving time.
The combined BATSE \& PVO distribution of burst intensities has
the `Euclidean' slope of $ -1.5 $ at the bright end, and a slope of
$ -0.8 $ at the faint end (Fenimore 1993), with the `roll-over'
caused by the cosmological redshift (Dermer 1992,
Mao \& Paczy\'nski 1992, Piran 1992).  According to Wijers (1997) the
following numbers follow from the `no evolution' scenario:
the energy per GRB is $ \sim 4 \times 10^{51} $ erg, the
energy generation rate is $ \sim 10^{53} ~ {\rm erg ~ Gpc^{-3} ~ yr^{-1}} $,
and the GRB rate per $ {\rm L^*} $ galaxy is 
$ \sim 4 \times 10^{-6} ~ {\rm yr^{-1}} $.

All these numbers were obtained assuming isotropic
GRB emission.  If the emission is beamed then the energy per burst
is reduced, the burst rate is increased by the same factor, but the GRB
energy generation rate per ${\rm Gpc^3}$ remains unchanged at a value
$ \sim 5 \times 10^4 $ 
times lower than the rate at which supernovae generate kinetic energy in
their explosions.

Recently, the massive star formation rate was found to be $ \sim 10 $ times
higher at $ z \approx 1 $ than than it is at $ z = 0 $ (Lilly et al. 1996, 
Madau et al. 1996).  If the GRB rate follows the massive star
formation rate (Totani 1997) then the
the consequences are dramatic, as emphasized by Sahu et al. (1997) and
by Wijers et al. (1997).  The increase in the comoving GRB rate with the 
cosmological distance compensates various redshift effects responsible for
the `roll-over' in the counts, and extends the range of distances
over which the `Euclidean' slope of the counts holds.
As a result the distance scale to the bursts
is increased compared to the no evolution model.  According to Wijers (1997)
the energy per burst in this `evolutionary' scenario increases to
$ \sim 10^{53} $ erg, the rate of energy generation per galaxy is
reduced by approximately one order of magnitude, and the GRB rate
is reduced as well.  This makes the `evolutionary' GRBs more powerful
and less common than the `no evolution' bursts used to be. 
If the star formation
rate increases beyond the redshift $ z \approx 1 $, then
the distance scale to the bursts increase even more, making them
even more energetic and even less common.

It is too early to decide which of the several cosmological
distance scales is correct, but
with a few dozen GRB/afterglow redshifts the choice
will be clear.  In any case GRBs are very rare
compared to ordinary supernovae.

%\section 4
\section{A micro-quasar}

If GRBs are associated with star forming regions, and if hypernovae
are somehow related to supernovae, i.e. they are violent ends
of massive star evolution, then a microquasar scenario (Paczy\'nski 1993)
is a plausible explanation.

At the end of its nuclear evolution the inner iron core of a very massive 
star collapses into a few solar mass black hole.
We know this is a real process as about ten binary 
stars are known to have black hole components of $ \sim 10 ~ M_{\odot} $ 
(cf. Tanaka \& Shibazaki 1996, p. 615).
If the star is spinning rapidly then its angular momentum prevents all
matter from going down the drain, and a
rotating, very dense torus forms around the rapidly spinning Kerr black 
hole (Woosley 1993a).  The largest energy reservoir, which may in
principle be accessed with a super-strong magnetic field (cf. Blandford
\& Znajek 1977), is the rotational energy of the black hole:
$$
{\rm E_{rot,max} \approx 5 \times 10^{54} ~ 
[erg] ~ \left( { M_{BH} \over 10 ~ M_{\odot} } \right) } .
\eqno(1)
$$
The maximum rate of energy extraction by the field was estimated by
Macdonald et al. (1986, eq. 4.50) to be
$$
{\rm L_{B,max} \approx 10^{51} ~ [erg ~ s^{-1}] ~ 
\left( { B \over 10^{15} ~ G } \right) ^2
\left( { M_{BH} \over 10 ~ M_{\odot} } \right) ^2 } .
\eqno(2)
$$

It is not clear how a superstrong field is generated, even though
it has become popular in theoretical papers over the last few years
Paczy\'nski 1991, 
Duncan \& Thompson 1992, Narayan, Paczy\'nski \& Piran 1992, Usov 1992,
Paczy\'nski 1993, Woosley 1993a,b, Hartmann \& Woosley 1995, Woosley 1995,
....., Vietri 1996, 
M\'esz\'aros \& Rees 1997, and many other).  The following is a possible
scenario.  A rapidly rotating massive star, just prior to its core collapse,
has a convective shell (Woosley 1997).  According
to Balbus (1997) a large scale magnetic field may be generated in the shell,
and it may reach equipartition with the convective kinetic energy density.
Following the collapse the polar caps of the shell end up in
the black hole, while the equatorial belt becomes part of the torus.
At least two different field topologies may emerge.  In one case the
magnetic field lines link the torus to the black hole, while in the
other case the field connection is severed.  In both cases
the collapse increases the field strength while the magnetic flux is
conserved, and a substantial radial component leads to a rapid field
increase driven by differential rotation.  If there is
no magnetic link between the torus and the black hole, then the magnetic field
helps to release gravitational energy associated with the torus accretion.
If a magnetic link is preserved then
a much larger rotational energy of the black hole can be extracted by
the Blandford \& Znajek (1977) mechanism, creating a microquasar.

It is well established that AGNs/blazars have relativistic jets which
generate strong and rapidly variable gamma-ray emission (cf. Ulrich et al.
1997, and references therein).  It is thought that the underlying `central
engine' is a supermassive black hole with a disk/torus of matter which 
provides accretion energy or the magnetic field confinement.  While theorists
argue about the specific mechanism in which blazars produce the observed
gamma-ray emission, the emission is there.  The formation of a similar
structure on a stellar mass scale, a Kerr black hole with a massive
disk/torus, is not speculative at all, as it is a natural end product
of massive star evolution.  The presence of a superstrong magnetic field,
and the ability of the system to generate a gamma-ray burst is just a
speculation at this time.  However, the observed properties of blazars
make this speculation plausible.

A pre-microquasar must be a member of a short period massive binary in
order to be rapidly rotating prior to core collapse.  Single stars lose
most of their angular momentum when they evolve to a red giant phase.
A member of a binary retains rapid rotation thanks to the
tidal interaction with the companion star.  The examples of such systems 
are the short period Wolf-Rayet binaries, and in 
particular Cyg X-3, with its $ \sim 5 $ hour orbit.

%\section 5
\section{Discussion}

A few different terms have been introduced in this paper in reference
to the objects which may be responsible for gamma-ray bursts.

The term hypernova is proposed to name the
phenomenon which is obviously explosive, and which is much more luminous
and energetic than any supernova.  Considering the energetics of the 
GRB/afterglow phenomenon, a term `hypernova'
seems more reasonable than `failed supernova' (Woosley 1993a) or 
`mini-supernova' (Blinnikov \& Postnov (1997).
It is likely that optical afterglows
unrelated to any GRBs will be detected in future massive variability
searches (Rhoads 1997); the term `hypernova' will be more
appropriate for such optical events than the `afterglow'.

The term `clean' fireball is often used to describe the popular model.
It is `clean' by design, to maximize the efficiency of conversion
of the kinetic energy into gamma-ray emission.  The rationale behind
this design is the perceived energy problem: how to obtain the 
$ \sim 10^{51} $ erg in gamma-rays out of merging neutron stars?
The energy problem may be even more acute if the new `evolutionary'
distance scale
turns out to be correct (Wijers et al. 1997).  The concept of a `dirty'
fireball is more natural, as any explosive event is likely to create
ejecta with a large range of specific kinetic energies or, in the
case of a relativistic explosion, a large range of Lorentz factors.
A fear of energy scale was never useful in astrophysics, as demonstrated
by the history of supernovae and quasars.
There is nothing in the laws of physics that would forbid explosions
with $ 10^{55} $ erg, or even more.  For those who are free of energy fobia
a `dirty' fireball appears more natural than a `clean' one.

Any dirty fireball model is likely to generate more or less
thermal emission from the optically thick, relatively slow
ejecta, at the very beginning of the explosion.  It is interesting
that Ginga experiment detected a number of X-ray precursors to
gamma-ray bursts.  In particular, the spectrum of X-ray precursor to
GRB 900126 was well fitted with a $ {\rm kT \approx 2 ~ keV} $
black body (Murakami et al. 1991).  The observed intensity corresponded
to the source radius of $ \sim 0.6 ~ {\rm km \times (d/1 ~ kpc) \approx
6 \times 10^{10} ~ cm \times (d/1 ~ Gpc) } $, where $ d $ is was the 
distance.  Recently, the presence of occasional X-ray precursors was 
reported by Sazonov et al. (1997).  

The concept of a `microquasar' (Paczy\'nski 1993,
Woosley 1993b, Hartmann \& Woosley 1995, Woosley 1995)
is introduced as a
specific example of a scenario in which a massive, rapidly rotating star
may generate over $ 10^{54} $ erg in kinetic energy of its ejecta upon
the end of its nuclear evolution.  While the geometry of the object,
a stellar mass Kerr black hole with a massive torus rotating around it,
is identical to the `failed supernova' scenario of Woosley (1993a),
the term `failed' does not seem appropriate for an event vastly more
energetic than any supernova.  As the neutrino driven explosion does
not appear to be feasible (Jaroszy\'nski 1996, 
Janka \& Ruffert 1996, Vietri 1996, Ruffert et al. 1997,
M\'esz\'aros \& Rees 1997, and references therein), a magnetically
driven event, analogous to the Blandford \& Znajek (1977) quasar model
is the next obvious candidate.  This may work if a superstrong
$ \sim 10^{15} $ G magnetic field is available to rapidly extract
the spin energy of the Kerr black hole and to use it to power a
relativistic explosion.

It is not likely that the concept of a GRB as a microquasar powered 
by the Blandford \& Znajek (1977) mechanism can be proven or
disproven on purely theoretical grounds.  It is useful
to realize, that while we have plenty of sound evidence
that Type II supernovae explode as a result of some `bounce',
or whatever process
following the formation of a hot neutron star, there is no
generally accepted physical process which would be efficient
enough to make this happen.  The theoretical problem with the SN II explosions
persists in spite of 2 or 3 decades of intense effort by a large
number researchers.  The problem is vastly worse with the GRBs as
they are $ 10^4 - 10^5 $ times less common than supernovae.  This
may imply that a very special set of circumstances is necessary
to generate the suitably energetic explosion.

While purely theoretical approach is difficult, some inferences
can be made without a quantitative model.
The death of a massive star cannot be more than a few million years
away from its birth time, and therefore it explodes within its star forming
region, or very close to it.  This makes it distinct from a
popular merging neutron star model: a merger follows orbital evolution
driven by gravitational radiation, long after the binary had formed.
During this time, $ \sim 10^8 - 10^9 $ years,
the system travels tens of kiloparsecs, having
acquired a high velocity during the two supernovae explosions (Tutukov
\& Yungelson 1994).

The star forming site for the GRBs in the microquasar scenario implies
that on many occasions the optical afterglow may be heavily obscured
by the dust commonly present in such regions (Jenkins 1997).
Gradual emergence of the fireball out of the
circum-stellar dust shell may affect the early 
afterglow, possibly accounting for the early rise in the
970508 optical light curve.

In the microquasar scenario the energy is released in a region
full of debris of the collapsing star.  Only a small fraction of 
all energy is likely to end up in the most relativistic ejecta,
which are responsible for gamma-ray emission following the standard
fireball scenario.  The bulk of kinetic energy is likely to be
associated with the much more
massive, and less relativistic ejecta.  In other words, a microquasar
is likely to create a dirty fireball.  This has an important consequence
for the afterglow.  In a clean fireball model the energy that powers
the afterglow is the residual kinetic energy of what is left of the
original GRB shell.  In a dirty fireball, when the fastest leading
shell is decelerated by the ambient medium, the
slower moving ejecta gradually catch up, and provide a long
lasting energy supply to the afterglow, much larger than the one
related to the GRB shell.  Therefore, the
afterglow may persist for much longer than predicted by the standard, clean
fireball model.  

I have presented
a weak case for a relation between GRBs and star forming
regions, based on the existing observations of the 970228, 970508, and
970828 bursts and their afterglows.  The case will be proven or
disproven when we shall have a few dozen afterglows.  If it is
established that the bursts are found in or near star forming 
regions, then the merging neutron star scenario will have to be
abandoned, and some supernova-like event, a violent death of a
massive star, will become a likely explanation for the origin of GRBs.
The microquasar scenario is a possible candidate for such an event.

The recent observations of X-ray spectra (Murakami 1997) offer yet
another important promise for the future.  With high enough spectral
resolution it will be possible to measure the redshift of the X-ray
source, or a lower limit to the redshift, even if no optical afterglow
is detected.  This is important as the afterglows are more common in X-rays
than in optical domain.

\acknowledgments{It is a great pleasure to
acknowledge many stimulating discussions and useful comments by S. Balbus,
J. Goodman, R. Chevalier, M. Rupen, S. van den Bergh, C. Thompson,
M. Vietri, E. Waxman, R. A. M. J. Wijers, S. E. Woosley, and many 
participants of the morning coffees at Peyton Hall.
This work was supported with the NSF grants AST--9313620 and AST--9530478.}  

%\newpage

%REFERENCES

\end{document}